# ELECTRIC FIELD MEASUREMENTS MADE IN SPACE
By F.S. Mozer and O.V. Agapitov
University of California, Berkeley, Space Sciences Laboratory

**ABSTRACT**
The operating principles of a DC and low frequency electric field detector are developed, after which, examples of earlier important electric field measurements are presented, including, the first observation of parallel electric fields in the auroral acceleration region, the first observation of time domain structures in space, the first experimental verification of symmetric magnetic field reconnection, the first observations of triggered ion acoustic waves, and oblique whistlers that directly accelerate electrons. Future possible improvements in the electric field measurement technique are described.

**KEY POINTS**
1. The operation of an electric field detector is described
2. Examples of important measurements made by electric field detectors are described.
3. Future possible improvements of the measurement technique are discussed.

## I  INTRODUCTION

The major goal of satellite studies in space plasmas is to understand the processes that accelerate and heat the plasma. From Maxwell's equations and Newton's second law, the only force capable of such acceleration and heating (neglecting weak forces such as gravity) is the electric field. Thus, electric field measurements should be a central component of all plasma physics space missions. However, this is not the case, as is evidenced by the lack of electric field detectors on the payloads described in the 2024 National Academy Decadal Survey [National Academy's, 2024] of space missions and instruments prioritized for the next decade. A possible explanation for the absence of electric field instruments on planned missions may be an insufficient understanding of the important science results achieved by them. In the space age, more than two dozen satellites equipped with electric field instruments have been flown and there are more than 2000 publications in the refereed literature on their measured results. The purposes of this article are to briefly review the history of these electric field measurements, the fundamental techniques developed for successful measurements, some major scientific achievements, and future improvements in the measuring technique. These electric field measurements include information on the amplitudes of electrostatic waves, the amplitudes of time domain structures, the phase velocities of electromagnetic waves, the wave-particle interactions that accelerate the plasma, the non-linear wave modes and effects, etc.

In the mid-twentieth century, DC and low frequency fields had not been measured although higher frequency electric field spectra were routinely measured by on-board electric field experiments [Gurnett publications, 2025]. This deficiency was overcome by the first successful DC and low frequency electric field measurement on a sounding rocket [Mozer and Bruston, 1968] and on a satellite [Mozer et al, 1977] by using spheres instead of cylinders as the sensing elements and by current biasing the detector to make it sensitive to electric fields and not to fluctuating currents. The importance of spherical rather than cylindrical electric field sensors stems from the fact that cylinders on a rotating satellite experience varying photoemission with the spin to produce a spurious signal at the same spacecraft rotation period as does the external electric field. The need for and explanation of biasing will next be described in terms of simplified Langmuir Probe theory.

**II   BIASING THE ELECTRIC FIELD DETECTOR**

Langmuir Probe theory provides the potential of a body with respect to the nearby plasma by requiring that, in equilibrium, the sum of all currents to the body be zero. For a body in shadow, Figure 1a provides a graphic illustration of this requirement with the potential of the body relative to the nearby plasma described along the X-axis and the current to the body given by the Y-axis. Suppose that the potential of the body is positive. Then the entire electron thermal current is attracted to become a negative current to the body, as illustrated in the figure, which includes focusing effects as the body becomes more positive. When the body is negative, the electron thermal flux with energies less than the potential of the body is reflected, so the electron thermal current reaching the body decreases as the potential of the body becomes more negative, with the current decreasing exponentially with a scale equal to the electron temperature in a simple model. With opposite signs, the same phenomena happen to the proton thermal flux, with the proton current about a factor of 40 less than the electron thermal current because of the heavier proton mass. In the absence of photoemission or bias current (and neglecting other, smaller currents) the voltage at which the two currents are equal in magnitude and opposite in sign is at the illustrated negative voltage location called the floating potential in Figure 1a.

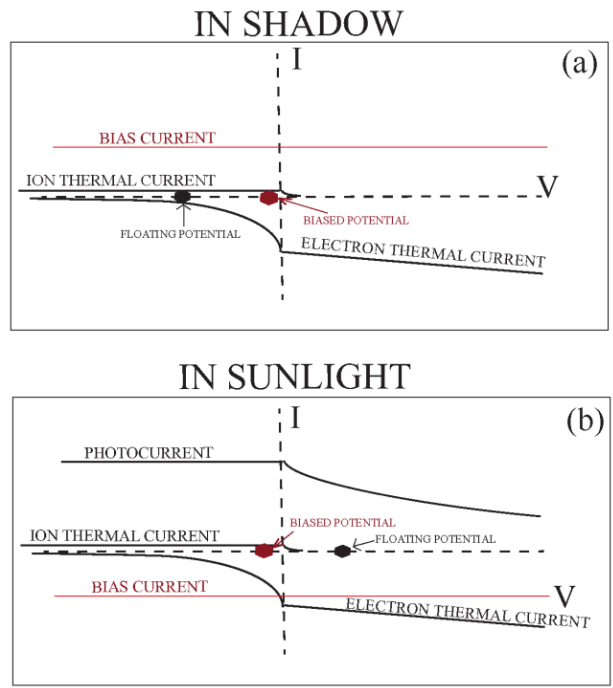

Figure 1.  Biasing the electric field detector in shadow (panel 1a) and in sunlight (panel 1b).

At this location, a small change of thermal current (or leakage current in the electronics, etc.) produces a large change of the floating potential, as can be seen in Figure 1a, so the potential is strongly dependent on current fluctuations. To resolve this problem, the potential of the body with respect to the nearby plasma should be nearly zero because, at this voltage, a change of thermal current produces a much smaller change of the potential of the body. Thus, a positive bias current, such as that illustrated in Figure 1a, makes the sum of all the currents equal to zero at the location labelled in the figure as the biased potential.

In sunlight, the photo current is generally larger than the electron thermal current, so the analysis requires inclusion of photoemission as in Figure 1b. For this case, the same analysis results in a negative bias current as shown in the figure.

An electric field instrument measures the potential difference between two separated surfaces, each of which is at the potential (floating or biased) of its surface. If the surfaces are unbiased, the floating potential of one surface may be greatly different from that of the other surface due to small differences of thermal current, secondary emission, leakage currents in the electronics, etc., so the observed potential difference is generally large and due to current imbalance and not electric fields. In this state it is not possible to make good measurements of the DC and low frequency electric fields, although, at higher frequencies, because the antennas are capacitively coupled to the measuring electronics [Fahleson, 1967], good wave measurements may be made.

In summary, for good electric field measurements, the spacecraft is at the floating potential and each antenna is biased to be at or near zero potential. The measured quantity, called the spacecraft potential, is then the difference between the biased potential of the antenna and the floating potential of the spacecraft. In this configuration, good electric field measurements are made as the difference between the spacecraft potentials of two separated spheres to produce the following examples of significant results for magnetospheric and heliospheric physics.

## III   THE AURORAL ACCELERATION REGION

In the 1960s it was found that electron precipitation in discreet auroral arcs consisted of nearly monoenergetic electrons. A natural mechanism for creating such beams is acceleration through electric fields that are parallel to the local magnetic field although such parallel electric fields were hard to explain theoretically. In 1976, the S3-3 satellite was launched to an apogee altitude of about 8000 km on auroral field lines and it carried the first current-biased, three component, spherical sensor, electric field measurement flown in space [Mozer et al, 1977]. In an early orbit it discovered ~500 mV/m parallel and perpendicular electric fields that are illustrated in the three panels of Figure 2. The parallel electric field (top panel of the figure) was huge and the perpendicular electric field in the bottom two panels of Figure 2 appeared as bipolar signatures, first of one sign then the other. This was interpreted by the potential structure illustrated in the image in Figure 2, in which the spacecraft, moving in latitude, flew mainly through the pair of oppositely directed perpendicular potentials with the largest part of the parallel electric field below the satellite altitude. This low altitude acceleration region is still far from understood and a modern satellite, with appropriate instrumentation, should be dedicated to auroral acceleration in order to understand the fields, waves, and plasma interactions in fuller detail.

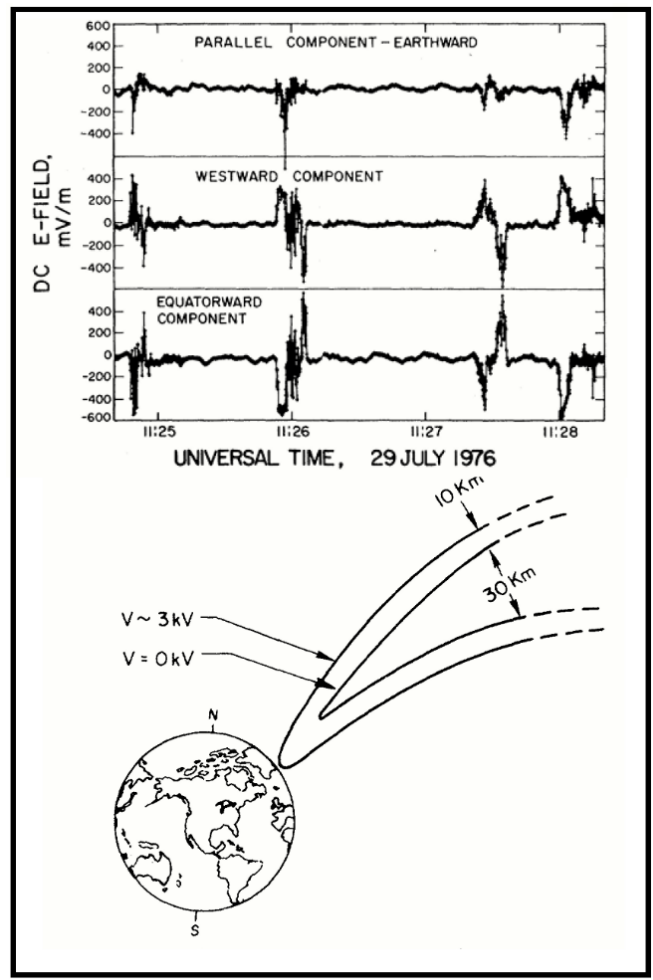

Figure 2. Parallel electric fields measured on the S3-3 satellite.

## IV  TIME DOMAIN STRUCTURES

Time domain structures (TDS) is the name given to ≥1 millisecond duration parallel electric field pulses. They are also called double layers, slow electron holes, ion holes, solitary waves, phase space holes, electron holes, etc. They are abundant through space and they accelerate electrons by processes such as Landau trapping, Fermi acceleration, and pitch angle scattering. Their properties are described in a review article [Hutchinson, 2024] that contains more than 250 references and in a summary article [Mozer et al, 2015].

To observe these structures, it is necessary to fly a current-biased, three-component, electric field experiment on a satellite. The first such example found in space is illustrated in Figure 3 [Temerin et al, 1982], which shows the three components of electric field for two events. In each event, the parallel electric field in the bottom panel of each display, shows the short duration spikes that are the time domain structures

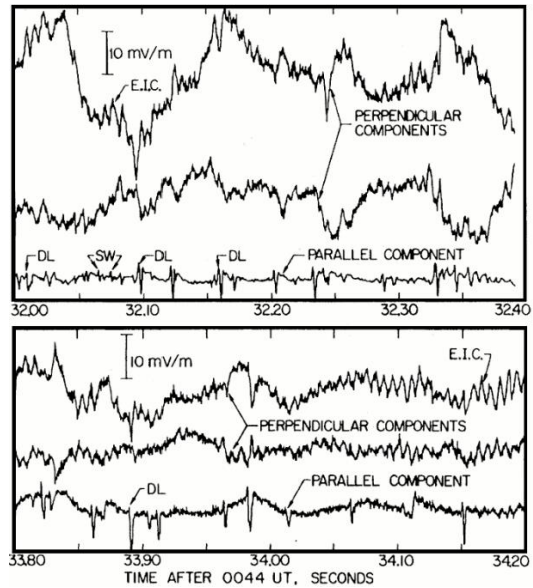

Figure 3. First example of time domain structures observed in space as seen in the parallel components of the electric field in the bottom panels of the two examples of three components of the electric field.

## V  MAGNETIC FIELD RECONNECTION

Particle-in-cell (PIC) simulations of symmetric magnetic field reconnection produced bipolar signatures in the electric and magnetic field before such measurements in space were attempted. On April 1, 2001, the Polar satellite crossed a subsolar magnetopause associated with antiparallel magnetic fields as shown in Figure 4e.

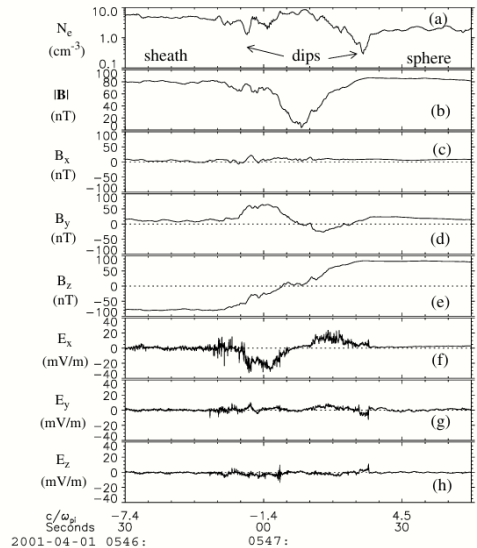

Figure 4. Electric and magnetic fields at a subsolar magnetospheric reconnection site. The bipolar magnetic and electric fields are seen in panels 4d and 4f.

Over a width of 6 magnetosheath ion skin depths, the magnetic field reversed (Figure 4e) and Hall magnetic (Figure 4d) and electric (Figure 4f) field signatures were observed [Mozer et al, 2002], which provided the first detailed experimental evidence for symmetric magnetic field reconnection. In addition to the Hall electric field, a reconnection event should contain small parallel electric fields that accelerate the electrons and that have not been observed to date. However, a large pulse of parallel electric field in a reconnection region has been observed [Ergun et al, 2016] which adds to the complexity of the process and that requires new measurements to find and understand all such parallel electric fields.

**VI  TRIGGERED ION ACOUSTIC WAVES AND ELECTRON HEATING**

Prior to the launch of the Parker Solar Probe, it was expected that the solar corona would contain whistler waves that would be a major contributor to the local electron heating. Contrary to this expectation, whistler waves were not observed sunward of about 30 solar radii [Cattell et al, 2021]. Instead, a new wave mode, called Triggered Ion Acoustic Waves, was discovered by the electric field sensor [Mozer et al, 2021, 2022, 2023].

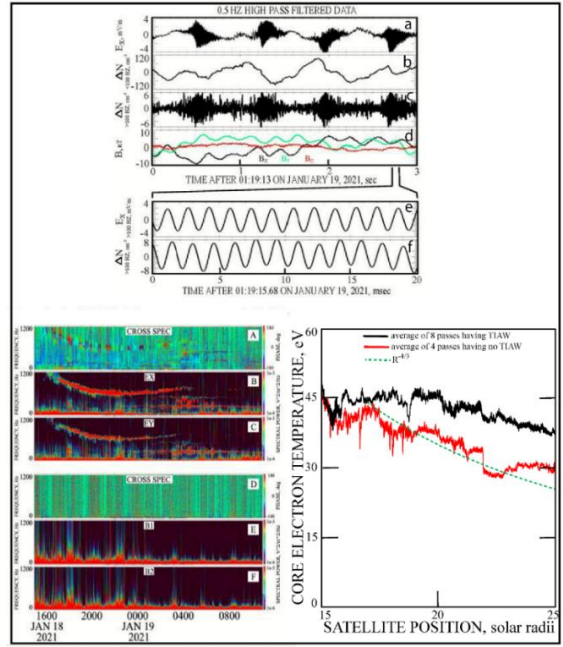

Figure 5. Triggered ion acoustic waves. The low and high frequency pair of synchronized waves are illustrated in Figure 5a, the pure tones of the electric field and density fluctuations are illustrated in Figures 5e and 5f, their many hour-long durations are illustrated in Figures 5B and 5C, and their heating of the electrons is shown by the electron spectra in the presence and absence of the triggered ion acoustic waves in the bottom right panel.

These waves have the following properties:
1. A pair of coupled waves, one of which is at a frequency of a few Hertz and the other is at few hundred Hertz (see Figure 5a).

2. The two waves are coupled such that the high frequency wave sometimes occurs in short bursts during successive periods of the low frequency sine wave and at a fixed phase of this lower frequency wave see Figures 5a).
3. The waves are electrostatic because they have no magnetic components (see Figure 5d) and they have density fluctuations at both frequencies (see Figures 5b and 5c).
4. Both the high and the low frequency waves are narrow band, effectively pure sine waves (see Figures 5e and 5f).
5. They can exist as a pair for times as long as several hours (see Figures 5B and 5C).
6. They are associated with heating the core electron distribution (see lower right-hand plot in Figure 5, which shows heating in the presence of triggered ion acoustic waves and no heating in their absence).

Thus, a new and exciting wave mode may be replacing whistler mode waves as the electron heat source near the Sun.

**VII   NON-LINEAR WHISTLERS AND THEIR ELECTRON ACCELERATION**

The discovery of large amplitude electric fields in whistler waves—exceeding 100 mV/m [Cattell et al., 2008; Cully et al., 2008; Kellogg et al., 2011; Wilson et al., 2012; Agapitov et al., 2014a] sparked renewed interest in nonlinear effects in wave-particle interactions. The interaction, specifically electron acceleration by coherent whistlers, was attributed to nonlinear cyclotron resonance, occurring when electrons move opposite to the direction of parallel whistler waves to satisfy resonance conditions, as well as Landau trapping [Artemyev et al., 2012, 2025]. Electric field data from the Electric Field and Waves (EFW) instruments onboard the Van Allen Probes A and B (August 30, 2012–May 2019) [Wygant et al., 2014]—a mission designed to study the radiation belts dynamics [Mauk et al., 2013] provided the first extensive evidence of nonlinear wave-particle interactions in space [Mozer et al., 2014, 2015, 2016b, 2018]. These interactions were associated with rapid electron acceleration (on the order of a bounce period) to tens or even hundreds of keV through Landau resonance in the parallel electric fields of time domain structures (TDS) [Artemyev et al., 2014a; Mozer et al., 2014; Vasko et al., 2015, 2017a,b] and chorus waves [Agapitov et al., 2015, 2018]. Observations revealed that most large-amplitude waves were highly oblique [Cattell et al., 2008; Agapitov et al., 2014, 2015] and exhibited electric fields aligned with the background magnetic field [Agapitov et al., 2014, 2015; Mozer et al., 2015]. A large electric field component parallel to the magnetic field enables interactions with electrons via Landau resonance, reducing the energy threshold for nonlinear trapping to ~1 keV [Artemyev et al., 2012, 2014b,c, 2016, 2025; Agapitov et al. 2014, 2015, 2018; Mozer et al. 2015; Drake et al. 2015]. This leads to dramatic changes in the particle distribution function over timescales of a few gyroperiods due to trapping [Agapitov et al., 2014a; Artemyev et al., 2012c]. Such interactions drive rapid acceleration (during one bounce period) along magnetic field lines, enhancing the "seed" population of 10–100 keV electrons [Artemyev et al., 2012; Agapitov et al., 2015]. Landau trapping also generates nonlinear feedback from trapped particles to the wave field, as recorded by EFW [Mozer et al., 2014], which may facilitate nonlinear wave-wave interactions [Mozer et al., 2015; Agapitov et al., 2015].

In Figures 6a and 6b, time domain waveforms of the magnetic and electric fields in a whistler wave measured during a 150 second interval are plotted. The electric field waveform is spiky because the wave is nonlinear. This point is emphasized in Figures 6c and 6d, which plot the power spectra of the two fields. The magnetic field spectrum shows only a single line while the electric field spectrum displays harmonics associated with the nonlinearity. A ten second example of the two waveforms in Figures 6e and 6f provides a better view of the nonlinearity in the electric field time domain waveform. The electron pitch angle distribution for different energies, displayed in Figure 6g, shows that several hundred electron volt to several kilovolt electrons had field aligned distributions that were produced by the nonlinear whistler wave. This illustrates the importance of nonlinear whistlers in the acceleration and pitch angle scattering of the local plasma [Agapitov et al. 2018].

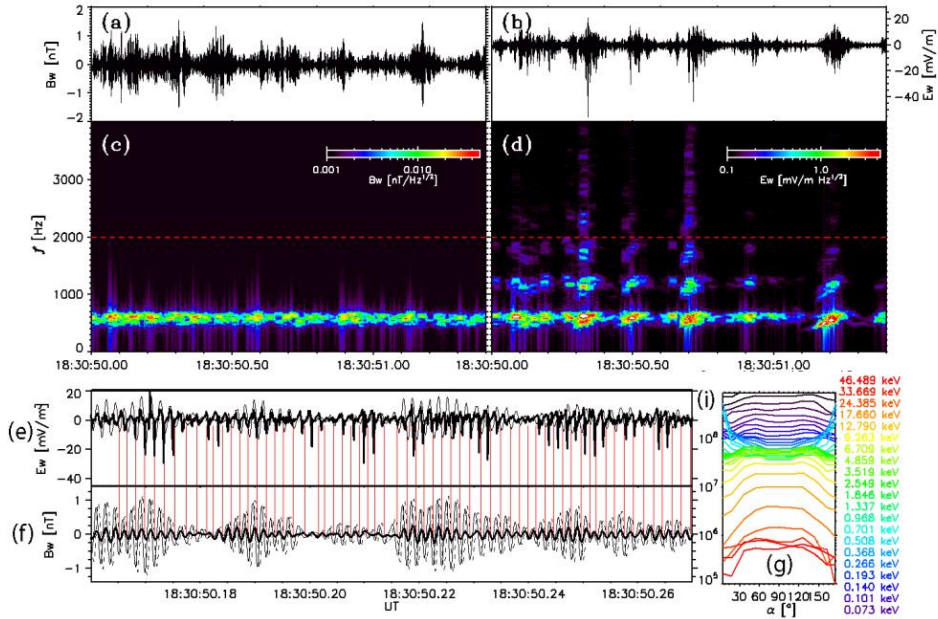

Figure 6. Description of a nonlinear whistler wave whose electric field spectrum (panel 6d) shows many harmonics of the fundamental wave that are produced by the nonlinearity at the same time that the electron pitch angle distribution of panel 6g shows field aligned hundreds of eV to several keV electrons accelerated by the nonlinear whistler wave [Agapitov et al., 2018].

**VIII   THE FUTURE**

Future space plasma missions should include electric field measurements to determine the modes and amplitudes of electrostatic waves, the amplitudes and observation rates of time domain structures, the phase velocities of electromagnetic waves, the wave-particle interactions that accelerate the plasma, the non-linear wave modes and effects, etc. Two such missions of high scientific interest should be measurement of the small parallel electric fields in magnetic field reconnection in order to understand particle acceleration in reconnection, and understanding the waves and processes that support the parallel electric field that causes discreet auroras.

Such measurements should be improved in at least the following ways:
1. An automated current biasing algorithm should be applied in order that the electric field detector can be in the best operating mode, independent of the plasma density and the presence or absence of sunlight. This algorithm should use either the on-board measured plasma or the value of the spacecraft potential as input for the determination of the optimum bias current. This algorithm should be tested on sounding rockets, as should a new geometric alignment of the electric field booms in order to measure a parallel electric field.

2. The typical data collection method of continuous measurement at low frequencies with an occasional, randomly timed, high frequency burst to look for waves, should be replaced by collection of all data at an exceedingly high rate for storage in an on-board many terabyte memory. This data could then be down sampled to produce a low frequency continuous waveform that is transmitted and scanned to determine the limited intervals for transmission of high-rate data. For example, if the three-axis electric field was measured at 1,000,000 samples/second, 20 days of continuous data could be stored in available, rad-hard, 10 terabyte solid state memories. The MMS satellite implemented a limited version of this data collection concept.

3. On typical spinning satellites, the two components of electric field in the spin plane have been well-measured because of the long booms and the separation of background offsets (which are DC signals) from the external electric field (which are AC signals at the spin rate). The third component of the field has been poorly measured because of the short booms along the spin direction and the mixing of the signal and the background offsets. The same statements may be made about magnetic field measurements. A new spacecraft design, called Grotifer, has been studied [Mao et al, 2015, Lejosne et al, 2022] to overcome these deficiencies by measuring all three components of both the electric and magnetic fields on rotating booms. This is achieved by having a central body fixed in inertial space with the body containing two rotating plates that are aligned at 90 degrees to each other and, on each of which, there is a two-axis measurement of the electric and the magnetic field such that all components of both fields are measured in rotating coordinate systems simultaneously. Recent lab simulations of this system have shown promise in achieving the control systems required to obtain and maintain this configuration.

# X ACKNOWLEDGEMENTS

The work of the many co-authors and engineers who produced this data is deeply appreciated. The work was funded at different times by the French Space Agency, the U.S. Office of Naval Research, and NASA. OVA was supported by NASA contracts 80NSSC22K0522, 80NSSC21K1770, and NASA's Living with a Star (LWS) program (contract 80NSSC20K0218). The work of FSM was supported by NASA contract NNN06AA01C.

OPEN RESEARCH

Data from the S3-3 satellite (Figs. 2 and 3) was obtained 50 years ago and is no longer available. However, this data is discussed in depth in Mozer et al, [1977]. Data from the Polar satellite (Fig. 4) may be obtained at https://cdaweb.gsfc.nasa.gov/, from the Parker Solar Probe (Fig. 5) at spdf.gsfc.nasa.gov, and from the Van Allen Probes (Fig 6) https://emfisis.physics.uiowa.edu/data/index and http://www.space.umn.edu/rbspefw-data/. Alternatively, all this data may be obtained from the NASA Space Physics Data Facility, https://spdf.gsfc.nasa.gov.